\documentclass[preprint]{aastex}

\begin{document}
\title{Chemical Composition of the Carbon-rich, Extremely Metal-Poor Star 
CS~29498--043: A New Class of Extremely Metal-Poor Stars with Excesses of 
Magnesium and Silicon\thanks{Based on data collected at Subaru Telescope, 
which is operated by the National Astronomical Observatory of Japan.}}

\author{Wako Aoki\altaffilmark{1}, John E. Norris\altaffilmark{2}, Sean G.
Ryan\altaffilmark{3}, Timothy C. Beers\altaffilmark{4}, Hiroyasu
Ando\altaffilmark{1}}
\altaffiltext{1}{National Astronomical Observatory, Mitaka, Tokyo, 181-8588 Japan; email: aoki.wako@nao.ac.jp, ando@optik.mtk.nao.ac.jp}
\altaffiltext{2}{Research School of Astronomy and Astrophysics, The Australian National University, Mount Stromlo Observatory, Cotter Road, Weston, ACT 2611, Australia; email: jen@mso.anu.edu.au}
\altaffiltext{3}{Department of Physics and Astronomy, The Open University, Walton Hall, Milton Keynes, MK7 6AA, UK; email: s.g.ryan@open.ac.uk}
\altaffiltext{4}{Department of Physics and Astronomy, Michigan State University, East Lansing, MI 48824-1116; email: beers@pa.msu.edu}

\begin{abstract} 

We analyze a high-resolution, high signal-to-noise spectrum of the carbon-rich,
extremely metal-poor star CS~29498--043, obtained with the Subaru Telescope
High Dispersion Spectrograph. We find its iron abundance is extremely low
([Fe/H] = $-3.7$), placing it among the few stars known with [Fe/H] $\le
-3.5$, while Mg and Si are significantly overabundant ([Mg/Fe] = +1.8, and
[Si/Fe] = +1.1) compared with stars of similar
metallicity without carbon excess.  Overabundances of N and Al were also
found. These characteristics are similar to the carbon-rich, extremely
metal-poor star CS~22949--037. Though the sample is small, our 
discovery of CS~29498--043 suggests the existence of a class of
extremely metal-poor stars with large excesses of C, N, Mg, and Si.

\end{abstract}


\section{Introduction}\label{sec:intro}

The most metal-poor stars in the Galactic halo are believed to contain
the material ejected from the first generations of stars.  To investigate the
nucleosynthetic yields of supernovae, the chemical compositions of a number of
extremely metal-poor stars have been studied recently \citep[e.g.,
][]{mcwilliam95, ryan96}. Some objects show 
distinct chemical characteristics considered to result from
nucleosynthesis in a single massive star and its supernova.

An extreme example is
CS~22949--037. \citet{mcwilliam95} found this object to be an
extremely metal-poor ([Fe/H] $\sim -4.0$)\footnote{[A/B] =
$\log(N_{\rm A}/N_{\rm B})- \log(N_{\rm A}/N_{\rm B})_{\odot}$, and
$\log \epsilon_{\rm A} = \log(N_{\rm A}/N_{\rm H})+12$ for elements A
and B.} giant with $\alpha$--element excesses. \citet{norris01}
confirmed the excesses of C, Mg and Si, compared with iron, and
discovered an extremely large enhancement of nitrogen ([N/Fe] =
+2.7). \citet{depagne02} found significantly large excesses of oxygen
and sodium ([O/Fe] = +2.0, and [Na/Fe] = +2.1). Comparisons with model
predictions for the yields of supernovae 
\citep[e.g., ][]{woosley95,fryer01,heg02} with
zero heavy elements have been made to explain its abundance
characteristics.

During our study of very metal-poor stars using the
University College London coud\'e \'echelle spectrograph (UCLES) at
the Anglo-Australian Telescope (AAT) (e.g., Norris et al. 1996), we
found that the carbon-rich object CS~29498--043 exhibited strong Mg
lines. To conduct a more detailed study, we
obtained a high-resolution spectrum with the High Dispersion Spectrograph
(HDS) of the Subaru Telescope \citep{noguchi02}. Our analysis shows that
CS~29498--043 is an extremely iron-deficient giant ([Fe/H] $=-3.7$) which,
contrary to most stars of similar
metallicity, exhibits large excesses of C, N, Mg, and Si compared with Fe.
These characteristics are similar to, but more extreme than, those of
CS~22949--037.

In this Letter we report on the composition of the carbon-rich,
extremely metal-poor star CS~29498--043. Its CH and CN bands are
as strong as another carbon-enhanced, extremely metal-poor star CS~22957--027
([Fe/H]$\sim -3.4$, Norris et al. 1997b; Bonifacio et al. 1998), which we also
analyzed using a spectrum obtained with HDS for comparison.

\section{Observation and Measurements}\label{sec:obs}

High-resolution spectra of CS~29498--043 and CS~22957--027 were
obtained with HDS in July 2001. They cover the
wavelength range 3550-5250~{\AA}, with resolving powers $R=$ 50,000
and $R=$ 60,000, respectively.  Data reduction was performed in the
standard way within the IRAF environment\footnote{IRAF is distributed
by the National Optical Astronomy Observatories, which is operated by
the Association of Universities for Research in Astronomy, Inc.  under
cooperative agreement with the National Science Foundation.}. For the
7200~sec and 4592~sec exposures for CS~29498--043 and CS~22957--027,
the detected photons number 1640 and 3780 per
0.013~{\AA} pixel at 4320~{\AA}, and 3260 and 5220 per 0.0155~{\AA}
pixel at 5180~{\AA}, respectively.

\begin{figure}
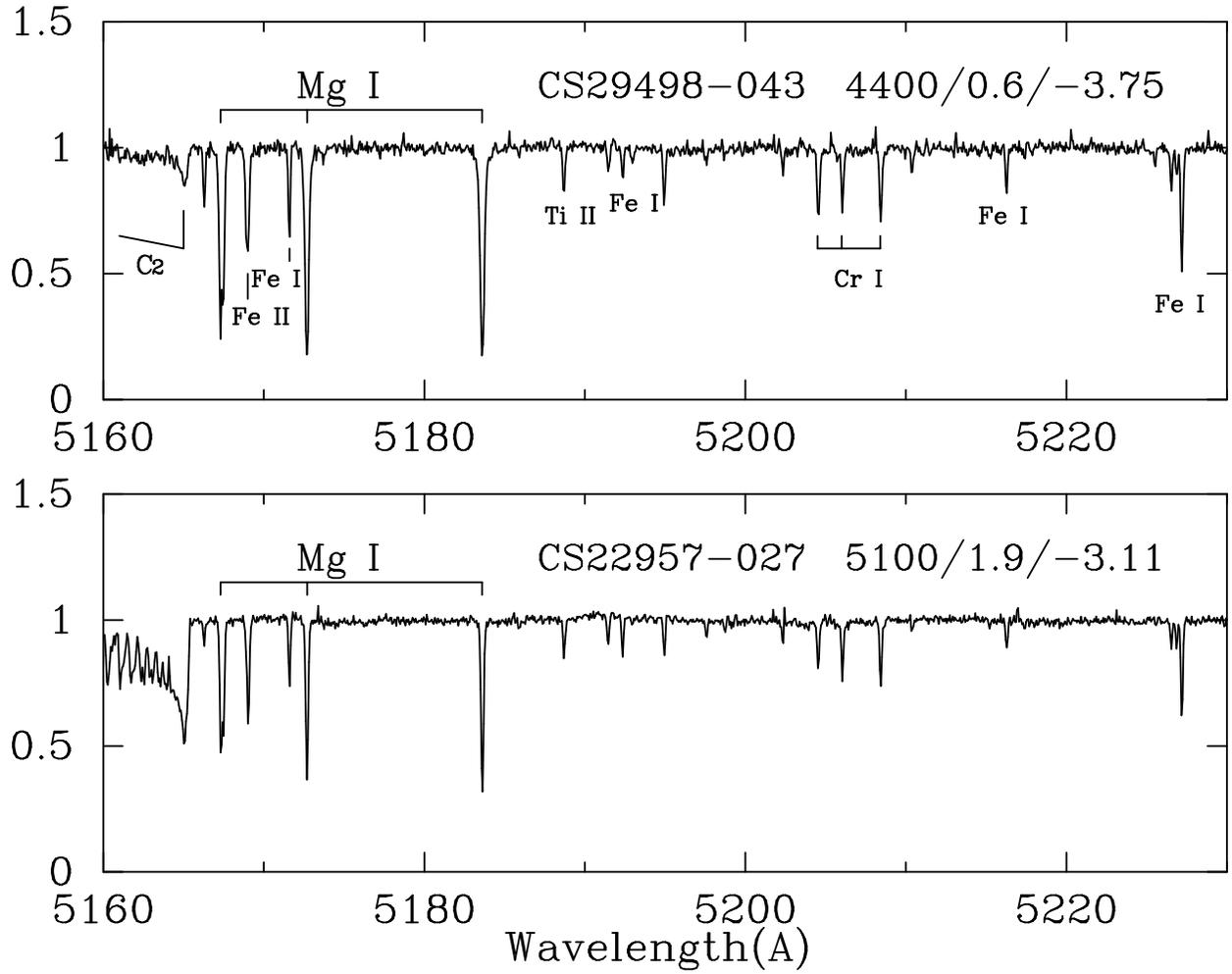

\caption[]{The observed spectrum of CS~29498--043 (upper panel) and that
of CS~22957--027 (lower panel). A five-pixel re-binning has been applied to
these spectra. The species which contribute to the absorption features are
shown. $T_{\rm eff}$, $\log g$, and [Fe/H] are also presented.  }
\label{fig:mg5170}
\end{figure}

The spectra around 5200~{\AA} are shown in Figure \ref{fig:mg5170},
where the Mg {\small I} triplet lines of CS~29498--043 are seen to be
much stronger than those in CS~22957--027, while other metallic
features, such as Fe {\small I} and Cr {\small I}, have similar
strengths in both spectra. This suggests a large enhancement of Mg
compared with Fe and other metals in CS~29498--043, though their
atmospheric parameters (e.g., effective temperature) must be
accounted for. The C$_{2}$ Swan band at 5165~{\AA} appears in both
spectra, indicating that both stars are carbon-rich.


Equivalent widths of absorption lines for 10 metals (12 ionization species) 
were measured by fitting Gaussian profiles.  
Wavelengths, excitation potentials, transition probabilities, and equivalent
widths will be reported separately (Aoki et al. 2002, in preparation).

Heliocentric radial velocities have been measured from a large number of
clean Fe {\small I} lines. The velocities of CS~29498--043 and
CS~22957--027 are $V_{\rm r}=-32.6 \pm 0.29$~km~s$^{-1}$ (26 July, 2001,
JD=2452117) and $V_{\rm r}=-61.6 \pm 0.22$~km~s$^{-1}$ (23 July, 2001,
JD=2452114), respectively. The radial velocity of CS~29498--043 was also
measured from the AAT/UCLES spectrum obtained on 14 Sept 2000 (JD = 2451802) to
be $V_{\rm r}=-32.5 \pm 0.18$~km~s$^{-1}$; there is no evidence of binarity.  
There is also no
indication of binarity for CS~22949--037 \citep{depagne02}. 
Further monitoring of radial velocities will provide a
valuable constraint on models of the formation of these objects.

\section{Abundance Analysis}

Effective temperatures ($T_{\rm eff}$) were estimated from the
$(B-V)_{0}$ and $(V-K)_{0}$ colors in Table \ref{tab:para}. The $B$
and $V$ photometry was from \citet{norris99} and \citet{beers92} for
CS~29498--043 and CS~22957--027, respectively, and $K$ was taken from
the interim 2MASS Point Source Catalogue
\citep{skrutskie97}. Interstellar reddening was estimated from the
\citet{schlegel98} maps.

The effective temperature
scale of carbon-rich stars by Aoki et al. (2002a) implied 
CS~29498--043 to be 4600~K from $(B-V)_{0}$, while 
4500~K is derived from $(V-K)_{0}$ using the scale
of \citet{alonso99}. Initially, we adopted $T_{\rm eff}=$ 4600~K for the
analysis of the Fe {\small I} and Fe {\small II} lines.  However, after
determining the other atmospheric parameters we obtained
lower abundances from higher excitation lines, suggesting that 
$T_{\rm eff}=$4600~K may be too high. Ultimately we
adopted $T_{\rm eff}$ = 4400~K,
but considering the ambiguity, we also performed analyses
for $T_{\rm eff}$=4600~K (from $(B-V)_{0}$) and for 
$T_{\rm eff}$=4250~K (at which the dependence of the derived Fe
abundance on excitation potential almost vanishes).

The effective temperature of CS~22957--027 estimated from $(B-V)_{0}$
is 5100~K, which agrees with the temperature
from $(V-K)_{0}$. The effective temperature 
derived from $(B-V)_{0}$ by \citet{norris97b} and \citet{bonifacio98}
is about 4850~K, but the effect of molecular absorption (e.g., CH
bands) in the blue range was not included in those works. The offset estimated
from the scale of Aoki et al. (2002a) is about
200~K, and is the likely reason for the discrepancy between the values.
\citet{norris97b} also performed an analysis for $T_{\rm
eff}$=5100~K as an alternative possibility, taking account of the
carbon excess. CS~22957--027 was also
studied by \citet{preston01}, who 
included the effect of molecular absorption on the broadband colors and
adopted $T_{\rm eff}=$5050~K.
Their value agrees well with that adopted here.

Using model atmospheres from \citet{kurucz93} at the adopted effective
temperatures, we performed abundance analyses in the standard manner
for the measured equivalent widths.  Surface gravities ($g$) were
determined from the ionization balance between Fe {\small I} and Fe
{\small II}, and the micro-turbulence ($v_{\rm tur}$) was determined
from the Fe {\small I} lines by demanding no dependence of the derived
abundance on equivalent widths.

We found that the Fe abundance of CS~29498--043 is very low ([Fe/H]
$\lesssim -3.5$, i.e., $\log \epsilon$(Fe)$\lesssim 4.0$), while the Mg
overabundance is quite large ([Mg/Fe] $\gtrsim 1.5$, i.e., $\log
\epsilon$(Mg)$\gtrsim 5.5$).
This means that Mg is an important electron source in the line-forming
layer of the stellar atmosphere, and the effect of its
overabundance must be included in the analysis. We iterated the
abundance analysis so that the derived Mg abundance is consistent with
the one assumed in the calculation. The effect of the Mg excess on the
determination of the surface gravity is large ($\Delta \log
g$=0.4~dex); if a higher Mg abundance is assumed, the
electron pressure is higher, and the ionization balance between Fe
{\small I} and Fe {\small II} demands a lower gravity. 
After taking into account a re-determination of the surface gravity, 
the effect of the Mg overabundance ([Mg/Fe] = +1.8) on the final
abundances of the elements examined here is not significant (e.g.,
$\Delta$[Fe/H] $<0.1$ dex). 
A high overabundance of Si ([Si/Fe] = +1.1) was also found, but its effect on
the analysis is negligible.

Spectrum synthesis was applied to the C$_{2}$, CN,
$^{12}$CH, and $^{13}$CH bands to determine the carbon and nitrogen
abundances and the carbon isotope ratio. Molecular data
were from \citet{aoki02c}.  For CS~29498--043, we tried the
analysis for the range $0.5\leq$ [O/Fe] $\leq 2.5$, mindful of
the large oxygen excess found in CS~22949--037 \citep{depagne02},
which may also apply to the present star.  We found that the carbon
abundance of CS~29498--043 derived from the C$_{2}$ Swan 0--0 band is
[$^{12}$C/Fe] = +1.9 when [O/Fe]=2.0 is assumed, and changes
by about 0.2~dex over the assumed range of [O/Fe]. We have adopted
these results for the carbon abundance and its error due to the
uncertainty of the oxygen abundance.  The nitrogen abundance of
CS~29498--043 was determined from the 0--1 band of the 4215~{\AA} CN violet
system to be [N/Fe] = +2.3.  A similar analysis was
applied to CS~22957--027; [$^{12}$C/Fe] = +2.4 and [N/Fe] = +1.6
were derived, and found to be insensitive to the assumed oxygen abundance.

Carbon isotope ratios are estimated from the three CH $B-X$ lines
around 4000~{\AA}. Line positions were calculated using the molecular
constants of \citet{kepa96}. The wavelengths of the
$^{13}$CH lines were adjusted by fitting the synthetic spectrum to the
observed one for CS~22957--027, and were then applied to 
CS~29498--043. The $^{12}$C/$^{13}$C ratio estimated
for CS~22957--027 is 8 $\pm 2$, which agrees reasonably with the value
$^{12}$C/$^{13}$C = 10 derived by \citet{norris97b}. The ratio
$^{12}$C/$^{13}$C = 6$\pm 2$ was derived for CS~29498--043. Though
this ratio is formally lower than that of CS~22957--027, we cannot
insist on the reality of the difference. Also, we cannot conclude that
the $^{12}$C/$^{13}$C of CS~29498--043 is higher than the 
equilibrium value of the CNO cycle ($\sim 4$), while that of
CS~22957--027 is obviously higher than the equilibrium value.

A standard analysis was performed for most elements using 
measured equivalent widths. Hyperfine splitting and isotope shifts were
included in the analysis of \ion{Ba}{2} lines, using the line list of
\citet{mcwilliam98}, and adopting the isotope ratios of the solar-system
{\it r}-process component.  The effect on the derived abundance is
small ($\lesssim 0.1$~dex) because the Ba lines are weak in these stars. The
Al and Mn abundances were determined by spectrum synthesis from the Al {\small
I} 3961~{\AA} and Mn {\footnotesize} I 4030~{\AA} lines, because these lines
are contaminated by $^{13}$CH and $^{12}$CH lines ($B-X$ band), respectively.
Since these blends involve CH doublets,
the strength of each blending CH line can be estimated well from the 
other component. The results of the abundance analysis are given in
Table~\ref{tab:res}.

To estimate the random errors, we first calculated the
dispersion of the abundances for Fe {\footnotesize I} lines in each star.  We
assumed the random error in {\it gf}-values to be 0.1~dex, and added it in
quadrature to the above dispersion. The final random error in the mean adopted
abundance for each element was evaluated by dividing by $n^{1/2}$ (where $n$ is
the number of lines used in the analysis, given in Table~\ref{tab:res}).

Errors arising from uncertainties of
the atmospheric parameters were evaluated, for CS~22957--027, using
$\sigma (T_{\rm eff})=100$~K, $\sigma (\log g)=0.5$, and $\sigma
(v_{\rm tur})=0.5$~km s$^{-1}$.  For CS~29498--043, whose effective
temperature is more uncertain as noted above, we also performed the
abundance analyses assuming $T_{\rm eff}=4600$~K and $T_{\rm
eff}=4250$~K, instead of the temperature $T_{\rm eff}=4400$~K adopted
above. We re-determined the 
surface gravity, metallicity, and micro-turbulence, and derived the
elemental abundances for each effective temperature. We adopted the
differences of the abundances derived from the three analyses 
as the errors due to the uncertainty
of effective temperature for CS~29498--043. The errors 
arising from uncertainties
of its surface gravity and microturbulence were evaluated using
$\sigma (\log g)=0.5$ and $\sigma (v_{\rm tur})=0.5~$~km s$^{-1}$.

Finally, we derived the total uncertainty by adding in quadrature the
individual errors, and list them in Table \ref{tab:res}. The error due
to the uncertainty of the oxygen abundance assumed in the analysis is
included in the total uncertainty of the carbon abundance of
CS~29498--043. 

Corrections for NLTE are not included in the present analysis. However,
their effect is expected to
be systematic, and the following discussion, based on the relative
abundances between our objects and other metal-deficient stars, is not
significantly affected.

\section{Discussion and Concluding Remarks}\label{sec:disc}

\begin{figure}
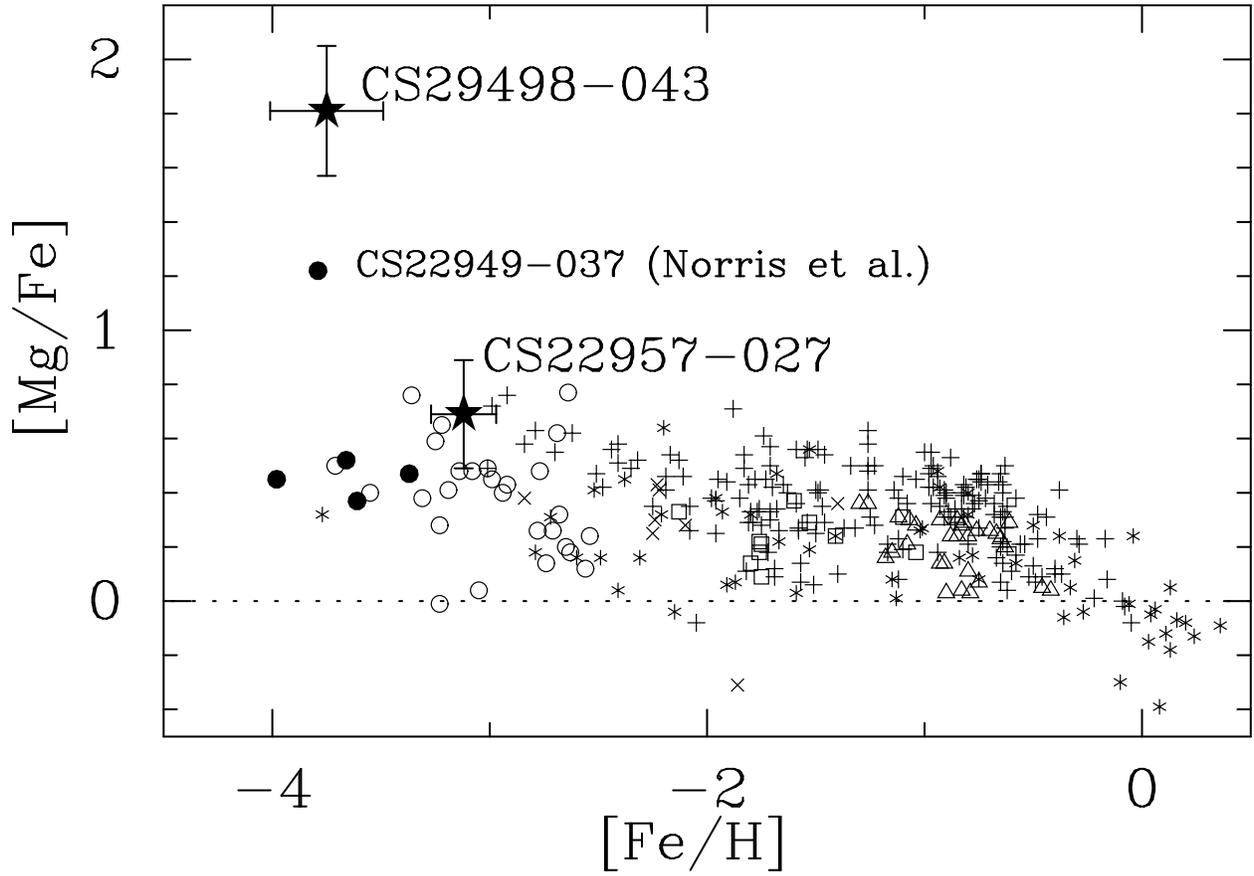

\caption[]{[Mg/Fe] as a function of [Fe/H]. The {\it stars} indicates
the values for CS~29498--043 and CS~22957--027 derived by the present
work.  The others are adopted from \citet[][{\it , filled
circles}]{norris01} and references therein.}
\label{fig:mgfe}
\end{figure}

Our analysis shows that CS~29498--043 is an extremely metal-poor
([Fe/H]$=-3.75$) star with a large excess of C, N, Mg, and Si. These
characteristics are similar to those of CS~22949--037
\citep{mcwilliam95, norris01, norris02}. In Figure \ref{fig:mgfe}, their
[Mg/Fe] values are shown as a function of [Fe/H], along 
with others from previous works compiled by Norris et al. (2001).
The [Mg/Fe] of CS~29498--043 is clearly much higher than the average for
other stars with similar [Fe/H], and is even higher than the already extreme
CS~22949--037. The [Mg/Fe] of CS~22957--027 follows the trend of the
stars with similar [Fe/H], as studied by \citet{norris97b} and
\citet{bonifacio98}.

Figure \ref{fig:abpat} shows the abundance differences
([X/Fe]$-<$[X/Fe]$>$) for CS~29498--043 and CS~22949--037 (Norris et
al. 2001) relative to the average values of the four stars
CD$-24^{\circ}$17504, CD$-38^{\circ}$254, CS~22172-002, and
CS~22885-096 studied by \citet{norris01}, as a function of atomic
number. Excesses of C, N, Mg, and Si clearly appear in both
CS~29498--043 and CS~22949--037, while there is no clear evidence of
departure of the relative abundances from zero for Sc-Ni. This
suggests that similar nucleosynthesis processes contributed to the
abundance patterns of these two stars.  The high [Mg/Fe] and [Si/Fe]
values presumably indicate that the heavy elements from which these
two stars formed came from supernovae whose outer layers (where C and
Mg are produced) escaped, but where relatively little material escaped
from nearer the iron core. Clearly, the nucleosynthesis mechanism is
distinct from that in the Asymptotic Giant Branch (AGB) stars
responsible for high C and s-process abundances in some, but not all,
carbon-enhanced, metal-poor stars.

While the excesses of Si in CS~29498--043 and CS~22957--027 are similar, the
excess of Mg in CS~29498--043 is noticeably larger than that of CS~22949--037;
the difference in [Mg/Fe] is
0.59~dex, significant at a 2.1-sigma level when we adopt the results
of \citet{norris01} for CS~22949--037. (The significance is 1.8-sigma if we
adopt the result of Depagne et al. 2002.) The Al abundance of CS~29498--043 is
also higher than that of CS~22949--037, although the uncertainty of the Al
abundance is large due to the contamination from the CH line and the strong
(but uncertain) NLTE effect for this element \citep[e.g., ][]{baumuller97}. In
addition to the small excess of the relative Al abundance, a large
enhancement of Na was also found in CS~22949--037
\citep{depagne02}. Studies of odd-$Z$ elements like Na and K in CS~29498--043
are desirable to investigate the nucleosynthesis processes which produced its
abundance pattern.  Measurement of its oxygen abundance, which has a large
excess in CS~22949--037 \citep{depagne02}, is also of importance.

The fraction of stars which are carbon-enhanced increases
with decreasing metallicity \citep{rossi99}. High-resolution
spectroscopy has been carried out for more than 10 carbon-rich stars
with $-3.0 < ${\rm [Fe/H]}$ < -2.0$. While some show large
excesses of $s$-process elements \citep[e.g., ][]{aoki02c}, others
show no such excess \citep{aoki02b}. No other carbon-rich object with
such large excesses of Mg and Si as CS~29498--043 is known
(Aoki, et al. 2002, in preparation).  However, only 4 objects with
[C/Fe]$\gtrsim 1.0$ are known in the metallicity range of [Fe/H]
$<-3.0$. One is the carbon- and nitrogen-enhanced star CS~22957--027
studied by \citet{norris97b}, \citet{bonifacio98}, and in the present
work. Another is the $r$-process-enhanced star CS~22892--052
\citep{sneden96}. The excesses of carbon and nitrogen of that object
([C/Fe]$\sim$[N/Fe]$\sim$1.0) are much smaller than those of
CS~22957--027, while the abundance pattern of the elements with
$12\lesssim Z\lesssim 28$ is similar to those of the other objects
with normal carbon abundances (McWilliam et al. 1995; Norris et al.
1997a).

The other two known carbon-rich stars with [Fe/H]$\lesssim -3$ 
are CS~22949--037 and CS~29498--043. These objects have lower iron
abundances ([Fe/H] $< -3.5$) than CS~22957--027 and CS~22892--052
([Fe/H] $\sim -3.0$), and also have large excesses of Mg and Si, as
shown here. Though the sample is too small to
permit definitive conclusions, the similarity of CS~22949--037 and CS~29498--043
suggests that other extremely metal-poor ([Fe/H]$\lesssim -3.5$)
stars with carbon excess may also show a large enhancement of Mg and
Si. Since the ratio of
carbon-rich objects seems to increase with decreasing metallicity, a
number of objects similar to these two stars perhaps exist.  Further
spectroscopic studies of candidate extremely metal-poor stars with
strong CH and CN features are essential to investigate the
nucleosynthesis processes in zero-(or very low-) metallicity stars in
the early Galaxy.  Comparisons with theoretical predictions of yields
ejected from supernovae will be presented separately in a future
paper; here we emphasize the importance of the discovery of
CS~29498--043, which suggests the existence of a class of extremely
metal-poor star with excesses of C, N, Mg, and Si.

\acknowledgements

The authors are grateful for fruitful discussions of these results with Drs A.
Chieffi, A. Heger, M. Limongi, and S. Woosley.  T.C.B acknowledges partial
support for this work from grants AST 00-98549 and AST 00-98508, awarded by the
US National Science Foundation.

\clearpage
\begin{deluxetable}{llcccccccc}
\tablewidth{0pt}
\tablecaption{PHOTOMETRIC DATA AND STELLAR PARAMETERS \label{tab:para}}
\startdata
\tableline
\tableline
Object & $V$ & $B$ & $(B-V)_{0}$ & $K$ & $(V-K)_{0}$ & $T_{\rm eff}$ & $\log g$ & $v_{\rm micro}$ & [Fe/H]\\ 
\tableline
CS~29498--043 & 13.72 & 14.80 & 0.99 & 10.95 & 2.50 & 4400 & 0.6 & 2.3 & $-3.75$ \\ 
CS~22957--027 & 13.59 & 14.36 & 0.74 & 11.61 & 1.92 & 5100 & 1.9 & 1.4 & $-3.11$ \\ 
\tableline
\enddata
\end{deluxetable}

\begin{deluxetable}{cccccccccc}
\tablewidth{0pt}
\tablecaption{ABUNDANCE RESULTS \label{tab:res}}
\startdata
\tableline
\tableline
 & \multicolumn{4}{c}{CS~29498--043} & & \multicolumn{4}{c}{CS~22957--027} \\
\cline{2-5} \cline{7-10}
Element \hspace{2cm} & [X/Fe] & $\log\epsilon_{\rm el}$ & $n$ & $\sigma$ & & [X/Fe] & $\log\epsilon_{\rm el}$ & $n$ & $\sigma$ \\
\tableline
$^{12}$C (C$_{2}$)\dotfill & $+$1.90 & 6.70 &    & 0.29 & &  $+$2.37  & 7.80 &   & 0.24 \\ 
N (CN)         \dotfill & $+$2.28 & 6.50 &    & 0.40  & &  $+$1.62  & 6.45 &   & 0.35 \\ 
Mg I          \dotfill  & $+$1.81 & 5.64 &  5 & 0.24 & &  $+$0.69  & 5.15 & 3 & 0.20 \\
Al I          \dotfill  & $+$0.34 & 3.08 &  1 & 0.36 & &  $-$0.77  & 2.60 & 1 & 0.33 \\
Si I          \dotfill  & $+$1.07 & 4.88 &  1 & 0.17 & &    ...    &  ... & ... & ... \\
Ca I          \dotfill  & $+$0.11 & 2.71 &  1 & 0.19 & &  $+$0.14  & 3.37 & 1 & 0.17 \\
Sc II         \dotfill  & $+$0.13 &$-$0.52 &  2 & 0.34 & &    ...    &  ... & ... & ... \\
Ti I          \dotfill  & $+$0.12 & 1.31 &  3 & 0.10 & &  $+$0.30  & 2.12 & 3 & 0.11 \\
Ti II         \dotfill  & $+$0.32 & 1.51 &  8 & 0.30 & &  $+$0.41  & 2.23 & 6 & 0.20 \\
Cr I          \dotfill  & $-$0.32 & 1.62 &  2 & 0.10 & &  $-$0.21  & 2.36 & 2 & 0.11 \\
Mn I          \dotfill  & $-$0.68 & 1.1 ~ &  1 & 0.37 & &  $-$0.41  & 2.0 ~ & 1 & 0.48  \\
Fe I ([Fe/H]) \dotfill  & $-$3.75 & 3.75 & 30 & 0.26 & &  $-$3.12  & 4.38 &25 & 0.15 \\
Fe II ([Fe/H]) \dotfill & $-$3.75 & 3.75 &  4 & 0.33 & &  $-$3.11  & 4.39 & 4 & 0.24 \\
Ni I          \dotfill  & ...     & ...  & .. & ...  & &  $-$0.25  & 2.88 & 4 & 0.26 \\
Sr II \dotfill          & $-$0.35 &$-$1.18 &  1 & 0.35 & &  $-$0.56  &$-$0.76 & 1  & 0.40 \\
Ba II \dotfill          & $-$0.45 &$-$1.98 &  2 & 0.20 & &  $-$1.23  &$-$2.13 & 2  & 0.21  \\
$^{12}$C/$^{13}$C \dotfill & 6 $\pm 2$ &&&&& 8 $\pm 2$ &&& \\ 
\tableline
\enddata
\end{deluxetable}

\clearpage

\begin{figure}
\caption[]{Relative abundances of elements in CS~29498--043 and
CS~22949--037, with respect to the average values of the four stars
studied by \citet{norris01} other than CS~22949--037, as a function of
atomic number. The Nitrogen abundance of the reference stars is
assumed to be [N/Fe] = 0. The elemental abundances of CS~22949-037 is
adopted from \citet{norris02} for nitrogen and from \citet{norris01}
for others. }
\label{fig:abpat}
\end{figure}

\plotone{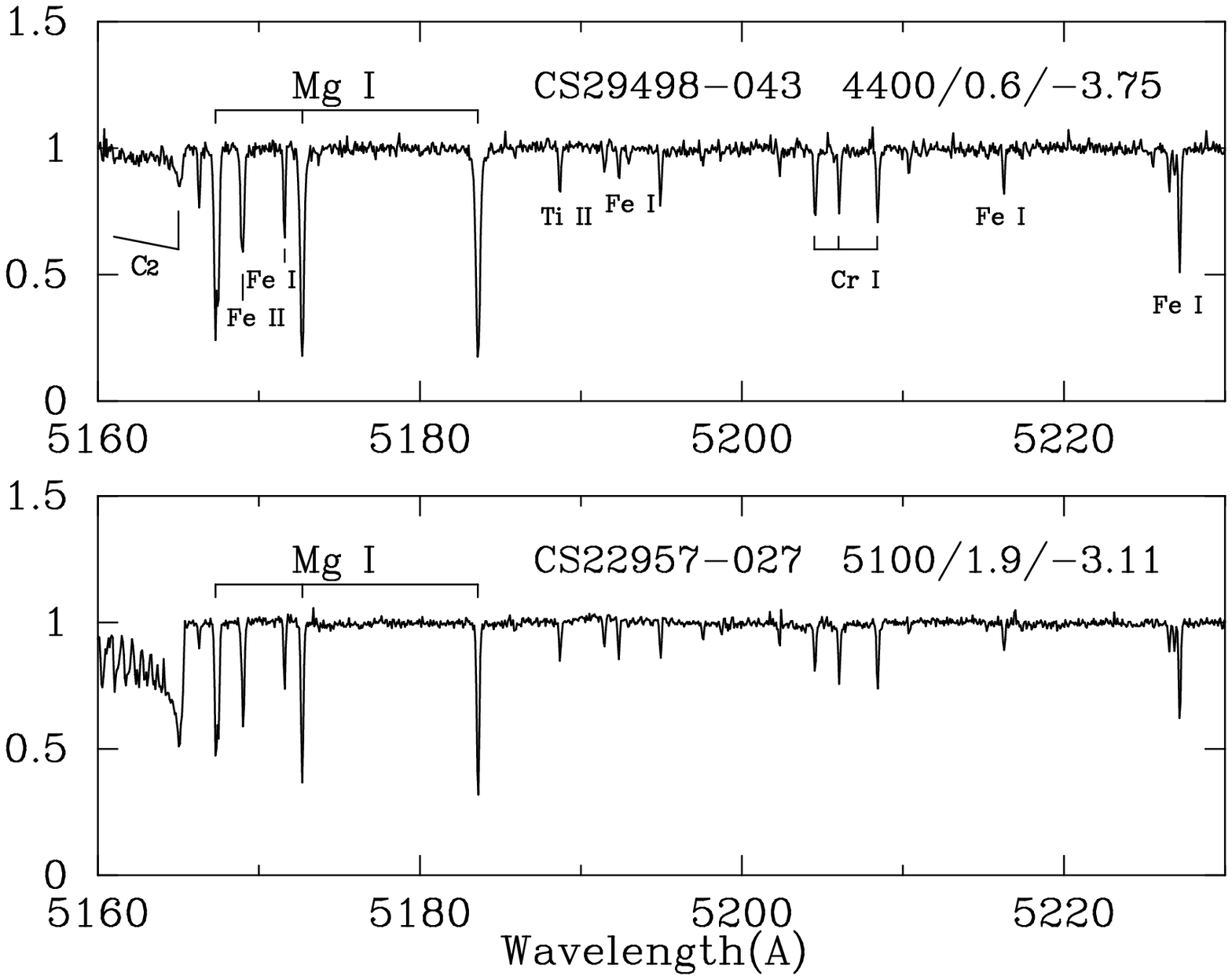}
\plotone{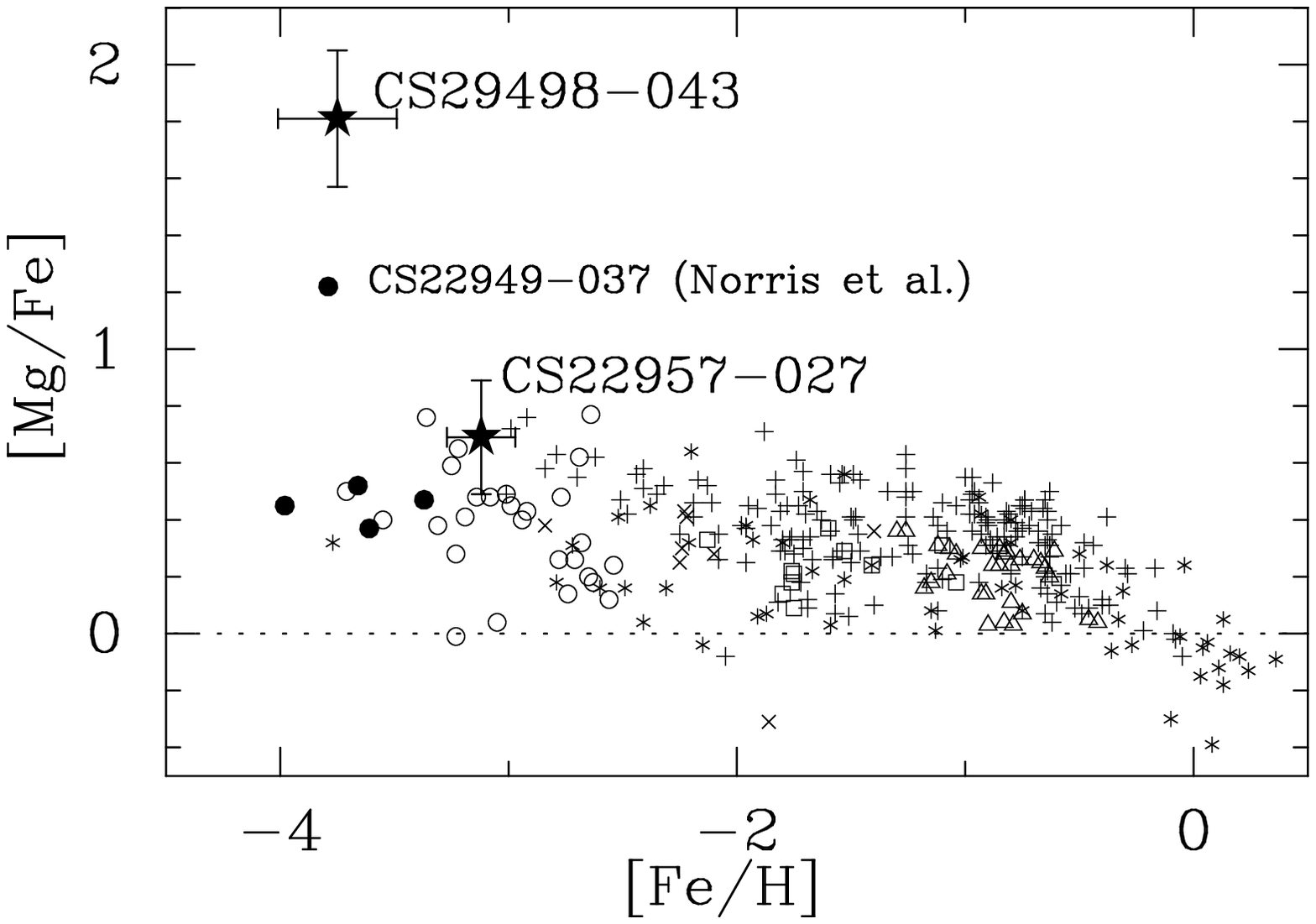}
\plotone{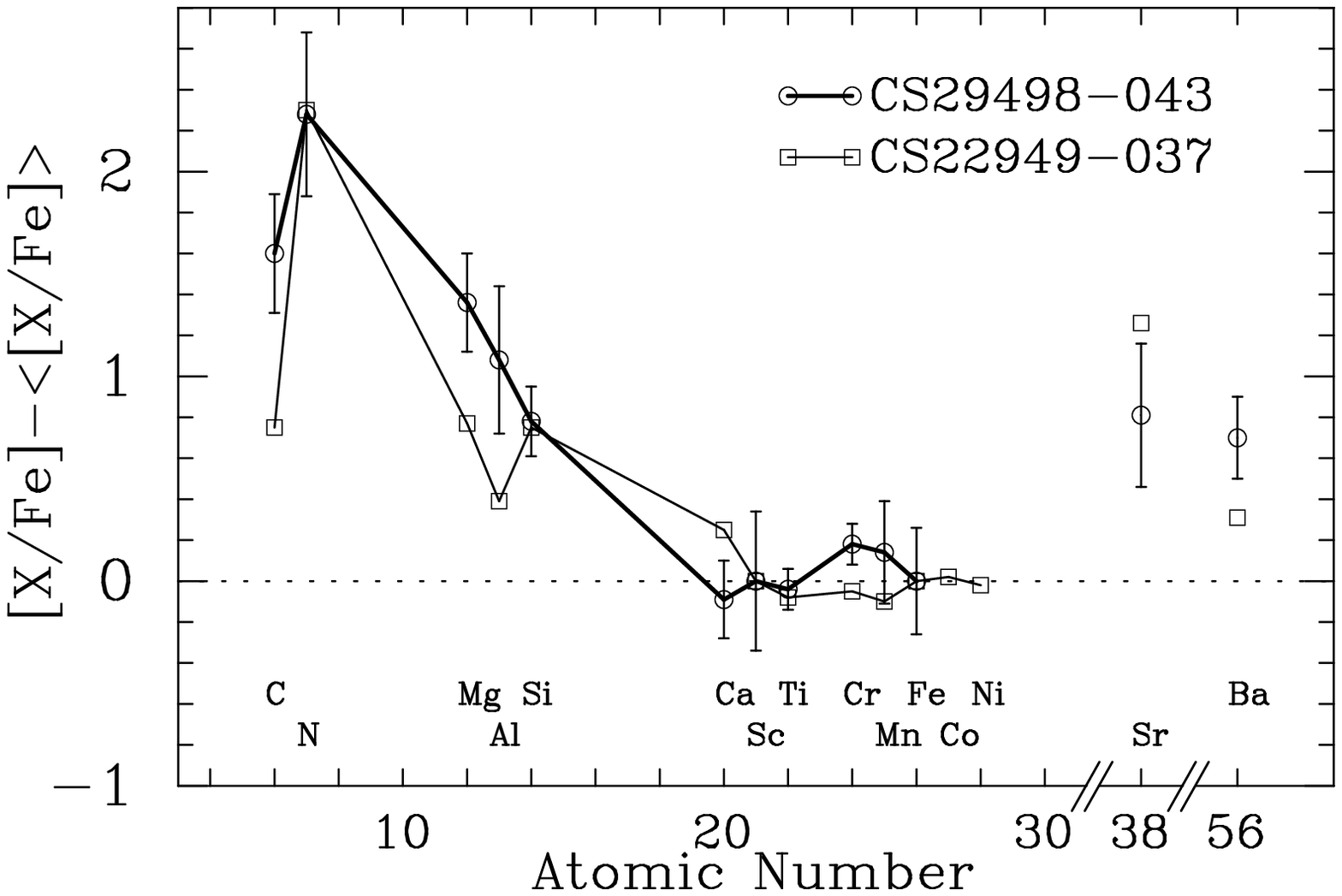}

\end{document}